\documentclass[10pt,conference]{IEEEtran}\IEEEoverridecommandlockouts
\usepackage{epsfig,graphics,subfigure,psfrag,amsmath,cases}
\usepackage{latexsym,amssymb,amsmath,epsfig,subfigure,algorithm,mathtools}
\usepackage{algorithmic}
\usepackage{times}
\usepackage{color}
\usepackage{url}

\usepackage{bm}

\usepackage{color}
\usepackage[makeroom]{cancel}

\author{Elena Boshkovska,
 Xiaoming Chen, Linglong Dai,   Derrick Wing Kwan Ng, and Robert Schober\thanks{ E. Boshkovska and R. Schober are with Friedrich-Alexander-University Erlangen-N\"urnberg (FAU), Germany. X. Chen is with Zhejiang University, Hangzhou, China. L. Dai is with Tsinghua University, Beijing, China. D. W. K. Ng is with The University of New South Wales, Australia. L. Dai is supported by the International Science \& Technology Cooperation Program of China (Grant No. 2015DFG12760) and the National Natural Science Foundation of China (Grant No. 61571270).
  R. Schober is supported by the AvH Professorship Program of the Alexander von Humboldt Foundation. D. W. K. Ng is supported under Australian Research Council's Discovery Early Career Researcher Award funding scheme (project
number DE170100137).}

}

\title{Max-min Fair Beamforming for SWIPT Systems with Non-linear EH Model}

\date{\thistime,\,\today}

\newtheorem{Thm}{Theorem}
\newtheorem{Lem}{Lemma}

\newcommand{\abs}[1]{\lvert#1\rvert}

\newcommand{\norm}[1]{\lVert#1\rVert}
\DeclareMathOperator{\Tr}{\mathrm{Tr}}
\DeclareMathOperator{\zero}{\mathbf{0}}
\DeclareMathOperator{\Rank}{\mathrm{Rank}}

\DeclareMathOperator{\vect}{\mathrm{vec}}
\DeclareMathOperator{\maxo}{\mathrm{maximize}}

\DeclareMathAlphabet\mathbfcal{OMS}{cmsy}{b}{n}

\newcommand{\qed}{\hfill \ensuremath{\blacksquare}}

\begin{document}
\IEEEspecialpapernotice{(Invited Paper)}
\maketitle

\begin{abstract}
We study  the beamforming design for  multiuser systems with simultaneous wireless
information and power transfer (SWIPT). Employing a practical
non-linear energy harvesting (EH) model, the design is formulated as a non-convex optimization problem  for the maximization of the minimum harvested power across several energy harvesting receivers. The proposed problem formulation takes into account   imperfect channel state information (CSI) and a
minimum required signal-to-interference-plus-noise ratio  (SINR). The globally optimal solution of the design problem is obtained via the semidefinite programming
(SDP) relaxation approach. Interestingly, we can show that at most one dedicated energy beam is needed to achieve optimality. Numerical results demonstrate that  with the proposed design a significant performance gain and improved fairness can be provided to the users compared to two baseline schemes.

\end{abstract}

\renewcommand{\baselinestretch}{0.955}
\large\normalsize

\section{Introduction}
In the past decades, the increasing interest in data hungry applications and heterogenous services has triggered
the  consumption of tremendous amounts of energy in wireless communication systems.  In practice,  mobile devices are
usually powered by batteries with  limited energy storage capacity which becomes a bottleneck in perpetuating the lifetime
of networks. To address this issue, energy harvesting (EH)-based communication technology has been proposed. In particular, this technology enables self-sustainability of power-constrained communication devices.
Communication systems may be equipped with energy harvesters \cite{Kwan:book_2017}\nocite{JR:imtiaz_hybrid}--\cite{JR:Kwan_hybrid_BS} to  scavenge energy from renewable natural energy sources such as solar and wind. Yet, these conventional energy sources are only available at specific locations which limits the mobility of portable devices. Besides,  the intermittent and uncontrollable nature of these natural energy sources is a concern for wireless communications, where uninterrupted and stable quality of service (QoS) are of paramount importance.

Recently, radio frequency (RF)-based wireless power transfer (WPT) has received considerable interest from both
industry and academia \cite{airfuel}\nocite{JR:SWIPT_mag,Krikidis2014}--\cite{Ding2014}. For example,  industrial companies  such as Samsung Electronics and Huawei Technology, have begun to launch research and study groups to facilitate the development and standardization of WPT \cite{airfuel}. In fact, the introduction of WPT avoids the high potential costs of planning, installing, displacing, and maintaining power cables in buildings and infrastructure. Specifically, for RF-based  communication networks, energy from ambient propagating electromagnetic (EM) waves in radio frequency (RF) can be harvested by energy-limited communication transceivers for prolonging their lifetimes and supporting the energy consumption required for future information transmission. This technology eliminates the need for power cords and manual recharging. Moreover, the broadcast nature of wireless channels facilitates one-to-many wireless charging and the possibility of simultaneous wireless information and power transfer (SWIPT) \cite{JR:SWIPT_mag}--\cite{Ding2014}. Compared to conventional
EH, RF-based EH technology can provide on-demand energy replenishment which makes it suitable
for smart wireless communication devices having strict QoS and energy
requirements.

In practice,  wireless power has to be transferred
via a signal with high carrier frequency such that antennas with small
size can be used for harvesting the power. However, the associated path loss severely
attenuates the signal leading to a small harvested power at the receiver.  Hence, multiple antenna beamforming has been proposed to facilitate efficient WPT \cite{JR:MIMO_WIPT}--\nocite{JR:rui_zhang,JR:Kwan_secure_imperfect,JR:Kwan_SEC_DAS}\cite{JR:Xiaoming_SWIPT_Massive}. In \cite{JR:MIMO_WIPT},  the concept of energy beamforming was first proposed to maximize the efficiency of WPT. In  \cite{JR:rui_zhang}--\cite{JR:Kwan_SEC_DAS}, energy beamforming was advocated to provide secure SWIPT in multiple-antenna systems. The authors of \cite{JR:Xiaoming_SWIPT_Massive} investigated  the impact of a massive number of antennas on the energy efficiency of  SWIPT systems.  However, most of the beamforming designs for SWIPT systems were based on an over-simplified linear EH model. In fact,  this model was recently shown  to be incapable of capturing the non-linear characteristics of practical RF EH circuits  \cite{JR:non_linear_model}. Besides, the results obtained in \cite{JR:MIMO_WIPT}--\cite{JR:Xiaoming_SWIPT_Massive} were based on the assumption of  perfect knowledge of the channel state information (CSI) of information receivers which is not realistic in practice.  Furthermore,  resource allocation  fairness was not considered in \cite{JR:MIMO_WIPT}--\cite{JR:Xiaoming_SWIPT_Massive} which may lead to an unsatisfactory performance for some users.

In this paper, we address these problems. In particular, we formulate the beamforming design as an optimization problem to provide max-min fairness in WPT to energy harvesting receivers equipped with practical non-linear energy harvesting circuits. The
optimization problem is solved by a semidefinite
programming (SDP) based resource allocation algorithm. Simulation results
illustrate an interesting trade-off between user fairness in energy harvesting
 and individual user data rate.

\textbf{Notation:} We use boldface capital and lower case letters to denote matrices and vectors, respectively. $\mathbf{A}^H$, $\Tr(\mathbf{A})$, and $\Rank(\mathbf{A})$ represent the Hermitian transpose, trace, and rank  of  matrix $\mathbf{A}$, respectively;  $\mathbf{A}\succ \zero$ and $\mathbf{A}\succeq \zero$ indicate that $\mathbf{A}$ is a positive definite and a  positive semidefinite matrix, respectively; $\mathbf{I}_N$ is the $N\times N$ identity matrix; $[\mathbf{B}]_{a:b,c:d}$  returns the $a$-th to the $b$-th rows and the $c$-th to the $d$-th column block submatrix of $\mathbf{B}$; $\mathbb{C}^{N\times M}$ denotes the set of all $N\times M$ matrices with complex entries; $\mathbb{H}^N$ denotes the set of all $N\times N$ Hermitian matrices.  The circularly symmetric complex Gaussian (CSCG) distribution is denoted by ${\cal CN}(\mathbf{m},\mathbf{\Sigma})$ with mean vector $\mathbf{m}$ and covariance matrix $\mathbf{\Sigma}$; $\sim$ means ``distributed as"; ${\cal E}\{\cdot\}$ denotes  statistical expectation; $\abs{\cdot}$ represents the absolute value of a complex scalar. $[x]^+$ stands for $\max\{0,x\}$, and   $[\cdot]^T$ represents the  transpose
operation.

 \begin{figure}[t]
 \centering
\includegraphics[width=3.5 in]{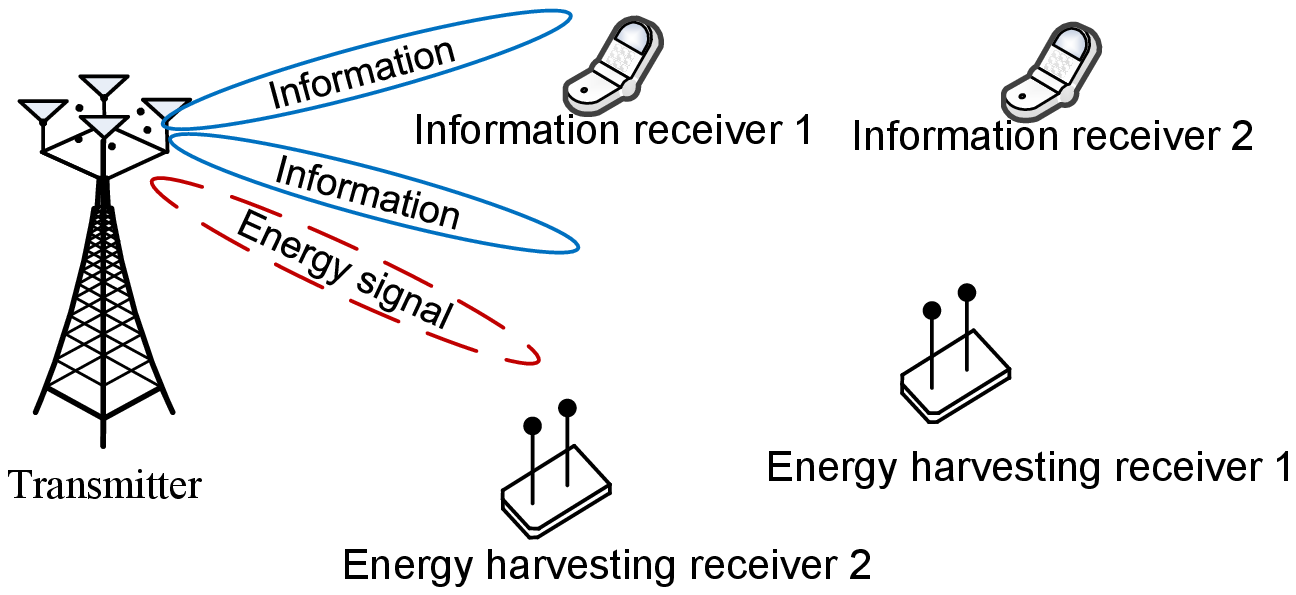}
 \caption{A  system model of a SWIPT system with $K=2$  IRs and $J = 2$ ERs.
 } \label{fig:system_model}
\end{figure}

\section{System Model}
In  this  section,  we  present  the   channel model  for downlink  SWIPT systems.

\subsection{Channel Model}\label{sect:model}
We consider a frequency flat and  slowly time varying downlink communication channel. In the SWIPT system, there are
  one transmitter, $K$  information receivers (IRs),  and $J$ EH receivers (ER), cf. Figure \ref{fig:system_model}. The transmitter  is equipped with $N_{\mathrm{T}}\geq 1$ antennas and serves both the IRs and the ERs simultaneously in the same frequency band.  We assume that each IR is a single-antenna device and each ER is equipped with $N_{\mathrm{R}}\geq 1$ receive antennas. The received signals at IR $k\in\{1,\ldots,K\}$ and  ER  $j\in\{1,\ldots, J\}$ are given by
\begin{eqnarray}
\hspace*{-4mm}y_k\hspace*{-2mm}&=&\hspace*{-2mm}\mathbf{h}^H_k\Big(\sum_{k=1}^K\mathbf{w}_ks_k+\mathbf{w}_\mathrm{E}\Big) +n,\,\, \mbox{and}\\
\hspace*{-4mm}\mathbf{y}_{\mathrm{ER}_j}\hspace*{-2mm}&=&\hspace*{-2mm}\mathbf{G}_j^H\Big(\sum_{k=1}^K\mathbf{w}_ks_k+\mathbf{w}_\mathrm{E}\Big)+\mathbf{n}_{\mathrm{ER}_j},\,\,  \hspace*{-1mm}\forall j\in\{1,\dots,J\},
\end{eqnarray}respectively, where  $\mathbf{w}_k\in\mathbb{C}^{N_{\mathrm{T}}\times1}$ and $s_k\in\mathbb{C}$ are the information beamforming vector and the associated information symbol for IR $k$, respectively.  Without loss of generality, we assume that ${\cal E}\{\abs{s_k}^2\}=1,\forall k$.  The vector channel of the transmitter-to-IR $k$ link is denoted by $\mathbf{h}_k\in\mathbb{C}^{N_{\mathrm{T}}\times1}$ and  the channel matrix  of the transmitter-to-ER  $j$ link  is denoted by $\mathbf{G}_j\in\mathbb{C}^{N_{\mathrm{T}}\times N_{\mathrm{R}}}$. Variables $n\sim{\cal CN}(0,\sigma_{\mathrm{s}}^2)$ and $\mathbf{n}_{\mathrm{ER}_j}\sim{\cal CN}(\zero,\sigma_{\mathrm{s}}^2\mathbf{I}_{N_{\mathrm{R}}})$  denote the additive white Gaussian noises (AWGN) at  IR $k$  and ER  $j$, respectively, where $\sigma_{\mathrm{s}}^2$ denotes the noise power at the receiver. $\mathbf{w}_\mathrm{E}\in\mathbb{C}^{N_{\mathrm{T}}\times 1}$ is a Gaussian pseudo-random sequence generated by the transmitter to facilitate efficient WPT.  In particular, $\mathbf{w}_\mathrm{E}$ is modeled as a complex Gaussian random vector with
 \begin{eqnarray}
  \mathbf{w}_\mathrm{E}\sim {\cal CN}(\mathbf{0},   \mathbf{W}_\mathrm{E}),
  \end{eqnarray}
where $\mathbf{W}_{\mathrm{E}}\in \mathbb{H}^{N_{\mathrm{T}}},\mathbf{W}_{\mathrm{E}}\succeq \mathbf{0}$, is the covariance matrix of the pseudo-random energy signal.
\subsection{Achievable Rate}
 Since the energy signal $\mathbf{w}_{\mathrm{E}}$ is a Gaussian pseudo-random  sequence which is known to all transceivers, IR $k$ can remove it via successive interference cancellation (SIC). Then, the achievable rate (bit/s/Hz) between the transmitter and IR $k$ is given by
\begin{eqnarray}
R_k&=&\log_2\Big(1+\frac{\mathbf{w}_k^H\mathbf{H}_k\mathbf{w}_k}
{ \sum_{i\ne k}\mathbf{w}_i^H\mathbf{H}_k\mathbf{w}_i+\sigma_{\mathrm{s}}^2}\Big),
\end{eqnarray}
where the interference caused by the energy signal, i.e., $\Tr(\mathbf{h}_k^H \mathbf{W}_{\mathrm{E}}\mathbf{h}_k)$ has been removed via SIC.

\subsection{Non-linear EH Model}
\begin{figure}[t]
 \centering
\includegraphics[width=3.5 in]{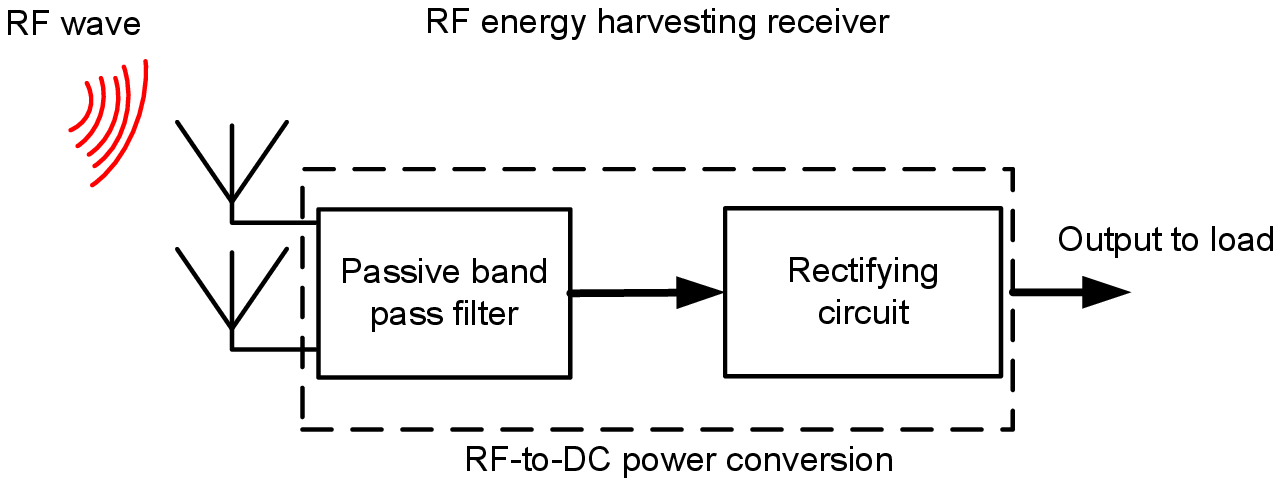}
 \caption{Block diagram of an ER. } \label{fig:circuit_model}
 \end{figure}

Figure \ref{fig:circuit_model} shows a general block diagram of an ER consisting of a passive filter and a rectifying circuit. In practice, the implementation of an ER depends on the adopted circuit components  which vary for different designs. To isolate the EH model from a specific circuit design, two general tractable  models, i.e., the linear model and the non-linear model,  have been proposed in the literature for characterizing the RF EH process. Mathematically, the total received RF power\footnote{In this paper, the unit of Joule-per-second is used for measuring energy. Thus, the terms ``power" and ``energy" are interchangeable.} at ER  $j$  is given by
  \begin{eqnarray}
P_{\mathrm{ER}_j}=\Tr\Big((\sum_{k=1}^K\mathbf{w}_k\mathbf{w}^H_k+\mathbf{W}_{\mathrm{E}}) \mathbf{G}_j\mathbf{G}_j^H\Big).
\end{eqnarray}

 For the linear EH model which as adopted e.g. in \cite{JR:MIMO_WIPT}--\cite{JR:Xiaoming_SWIPT_Massive}, the total harvested power at  ER  $j$, $\Phi_{\mathrm{ER}_j}^{\mathrm{Linear}}$, is typically modelled by the following linear equation:
\begin{eqnarray}\label{eqn:linear_EH_model}
\Phi_{\mathrm{ER}_j}^{\mathrm{Linear}}=\eta_j P_{\mathrm{ER}_j},
\end{eqnarray}\noindent
where $0\leq\eta_j\leq1$ is the constant power conversion efficiency of ER $j$.

 \noindent
Yet, practical RF-based EH circuits are inherently non-linear and the conventional linear model fails to capture this important characteristic, as shown experimentally in \cite{CN:EH_measurement_2}\nocite{JR:Energy_harvesting_circuit}--\cite{JR:EH_measurement_1}.   Motivated by this, a parametric non-linear EH model was proposed in \cite{JR:non_linear_model,7919062} which has been shown to be in excellent agreement  closely with practical measurement results. In this paper,  we adopt the more realistic  non-linear EH model and the total harvested power at ER  $j$, $\Phi_{\mathrm{ER}_j}$, is modelled as:
 \begin{eqnarray}\label{eqn:EH_non_linear}
\Phi_{\mathrm{ER}_j}\hspace*{-2mm}&=&\hspace*{-2mm}
 \frac{[\Psi_{\mathrm{ER}_j}
 - M_j\Omega_j]}{1-\Omega_j},\, \Omega_j=\frac{1}{1+\exp(a_jb_j)},\\
\mbox{where}\,\,\Psi_{\mathrm{ER}_j}\hspace*{-2mm}&=&\hspace*{-2mm} \frac{M_j}{1+\exp\Big(-a_j(P_{\mathrm{ER}_j}-b_j)\Big)}
  \end{eqnarray}
 is a sigmoidal function which has the received RF power, $P_{\mathrm{ER}_j}$,  as the input.  Three parameters,  namely,  $M_j$, $a_j$, and $b_j$, are introduced to describe the shape of the logistic function which depends on the physical properties of the RF EH circuit. Specifically,  $M_j$, $a_j$, and $b_j$ are constants which determine the maximum harvestable power,   the charging rate with respect to the input power, and the minimum required voltage to turn on the EH circuit\footnote{We note that these parameters can be determined for any given EH circuit. }, respectively.
%
\subsection{Channel State Information}
We take into account the imperfection of the channel state information (CSI) for beamforming design. To this end, we adopt a deterministic model \cite{JR:Kwan_SEC_DAS,JR:CSI-determinisitic-model2}. In particular,  we model the  CSI of the links as:
\begin{eqnarray}\label{eqn:outdated_CSI}
\mathbf{h}_k&=&\mathbf{\widehat h}_k + \Delta\mathbf{h}_k,\,   \\
{\bm\Upsilon }_k&\triangleq& \Big\{\Delta\mathbf{h}_k\in \mathbb{C}^{N_{\mathrm{T}}\times 1}  :\norm{\Delta\mathbf{h}_k}_2^2 \le \rho^2_k\Big\}, \label{eqn:outdated_CSI-set2}\\
\mathbf{G}_j&=&\mathbf{\widehat G}_j + \Delta\mathbf{G}_j,\,   \forall j\in\{1,\ldots,J\},  \mbox{   and}\label{eqn:outdated_CSI-set1}\\
{\bm\Xi }_j&\triangleq& \Big\{\Delta\mathbf{G}_j\in \mathbb{C}^{N_{\mathrm{T}}\times {N_{\mathrm{R}}}}  :\norm{\Delta\mathbf{G}_j}_\mathrm{F}^2 \le \upsilon_j^2\Big\},\forall j,\label{eqn:outdated_CSI-set}
\end{eqnarray}
respectively, where $\mathbf{\widehat h}_k$ and $\mathbf{\widehat G}_j$ are the estimates of channel vector $\mathbf{h}_k$ and channel matrix $\mathbf{G}_j$, respectively. The channel estimation errors of $\mathbf{h}_k$ and $\mathbf{G}_j$ are denoted by $\Delta\mathbf{h}_k$ and $\Delta\mathbf{G}_j$, respectively. Sets  ${\bm\Upsilon }_k$ and ${\bm\Xi }_j$  collect all possible channel estimation errors.  Constants $\rho_k$ and  $\upsilon_j$  denote the maximum values of the norms of the CSI estimation error vector   $ \Delta\mathbf{h}_k$ and the CSI estimation error matrix $ \Delta\mathbf{G}_j$, respectively.

%

\section{Problem Formulation and Solution}\label{sect:problem_formulation_solution}
The system objective is to maximize the minimum harvested power among all the ERs, i.e.,  to provide max-min fairness \cite{CN:PHY_SEC_max_min}, while guaranteeing the QoS of information communication. To this end, we formulate the resource allocation
algorithm design as the following non-convex optimization
problem:
\begin{eqnarray}\label{eqn:TP_maximization}
\underset{\mathbf{W}_\mathrm{E}\in \mathbb{H}^{N_{\mathrm{T}}},\mathbf{w}_k}{\maxo}\,\, && \hspace*{-3mm} \min_{\underset{j\in\{1,\ldots,J\}}{\Delta\mathbf{G}_j\in\mathbf{\Xi}_j}} \Psi_{\mathrm{ER}_j}\\
\mathrm{s.t.}\,\, &&\hspace*{-3mm}\mathrm{C1}:\,\,\sum_{k=1}^K\norm{\mathbf{w}_k}_2^2 + \Tr(\mathbf{W}_{\mathrm{E}})\leq P_{\max},\notag\\
&&\hspace*{-3mm}\mathrm{C2}:\,\,\min_{\Delta\mathbf{h}_k\in {\bm\Upsilon }_k} \frac{\mathbf{w}_k^H\mathbf{H}_k\mathbf{w}_k}
{ \sum_{i\ne k}\mathbf{w}_i^H\mathbf{H}_k\mathbf{w}_i+\sigma_{\mathrm{s}}^2} \geq \Gamma_{\mathrm{req}_k},\notag\\
&&\hspace*{-3mm} \mathrm{C3}:\,\, \mathbf{W}_{\mathrm{E}}\succeq \zero.\notag
\end{eqnarray}
\noindent
 The objective function in \eqref{eqn:TP_maximization} takes into account the CSI
uncertainty set $\mathbf{\Xi}_j$ to provide robustness against CSI imperfection.
 Constants $P_{\max}$ and $\Gamma_{\mathrm{req}_k}$  in constraints C1 and C2 denote the maximum transmit power allowance and the QoS requirement on the minimum received signal-to-interference-plus-noise ratio (SINR) at IR $k$, respectively.  Constraint C3 and $\mathbf{W}_{\mathrm{E}}\in \mathbb{H}^{N_{\mathrm{T}}}$ constrain matrix $\mathbf{W}_{\mathrm{E}}$  to be a positive semidefinite Hermitian matrix.

 It can be observed that constraint C2 is non-convex. Besides, there are infinitely many possibilities in both
the objective function and constraint C2, due to the CSI uncertainties. In order to design a computationally efficient resource allocation algorithm, we first define $\mathbf{W}_k=\mathbf{w}_k\mathbf{w}^H_k$ and  transform the considered problem into the following equivalent rank-constrained semidefinite program (SDP):

\begin{eqnarray}\notag
\underset{\underset{\bm{\beta},\tau}{     \mathbf{W}_\mathrm{E},\mathbf{W}_k\in \mathbb{H}^{N_{\mathrm{T}}},   }}{\maxo}\,\, \hspace*{-14mm}&& \hspace*{14mm}  \tau\\
\hspace*{-14mm} \mathrm{s.t.}\,\, &&\hspace*{-6mm}\mathrm{C1}:\,\,\sum_{k=1}^K\Tr(\mathbf{W}_k)+\Tr(\mathbf{W}_{\mathrm{E}}) \leq P_{\max},\notag\\
\hspace*{-14mm} &&\hspace*{-6mm}\mathrm{C2}:\,\, \min_{\Delta\mathbf{h}_k \in {\bm\Upsilon }_k } \frac{\Tr(\mathbf{W}_k\mathbf{H}_k)}{\Gamma_{\mathrm{req}_k}}\ge \sum_{i\neq k}\Tr(\mathbf{W}_i\mathbf{H}_k)+\sigma_{\mathrm{s}}^2,\notag\\
\hspace*{-14mm} &&\hspace*{-6mm}\mathrm{C3}:\,\, \mathbf{W}_{\mathrm{E}} \succeq \zero,\notag \\
\hspace*{-14mm} &&\hspace*{-6mm}\mathrm{C4}:\,\, \frac{M_j}{\Big(1\hspace*{-0.5mm}+\hspace*{-0.5mm}\exp\big(\hspace*{-0.5mm}-\hspace*{-0.5mm}a_j(\beta_j\hspace*{-0.5mm}-\hspace*{-0.5mm}b_j)\big)\Big)} \geq \tau(1\hspace*{-0.5mm}-\hspace*{-0.5mm}\Omega_j)\hspace*{-0.5mm}+\hspace*{-0.5mm}M_j\Omega_j,\hspace*{-0.5mm}\forall j,\notag\\
\hspace*{-14mm} &&\hspace*{-6mm}\mathrm{C5}:\,\, \min_{\Delta\mathbf{G}_j\in{\bm\Xi }_j} \Tr\Big((\sum_{k=1}^K\mathbf{W}_k+\mathbf{W}_{\mathrm{E}})\mathbf{G}_j\mathbf{G}_j^H\Big)\ge \beta_j,\forall j,\notag\\
\hspace*{-14mm} &&\hspace*{-6mm}\mathrm{C6}:\,\, \Rank(\mathbf{W}_k)\le 1,\notag\\
\hspace*{-14mm} &&\hspace*{-6mm}\mathrm{C7}:\,\, \mathbf{W}_k \succeq \zero,\label{eqn:Rank-constrained}
\end{eqnarray}
\noindent
where $
\mathbf{H}_k=\mathbf{h}_k\mathbf{h}^H_k$.  Vector $\bm{\beta}=\{\beta_1,\ldots,\beta_j,\ldots,\beta_J\}$ and $\tau$ are auxiliary optimization variables. We note that C7 and $\Rank(\mathbf{W}_k)\leq 1$ in (\ref{eqn:Rank-constrained}) are imposed such that $\mathbf{W}_k=\mathbf{w}_k\mathbf{w}^H_k$.
 Now, the transformed problem in \eqref{eqn:Rank-constrained} involves infinitely many constraints only in C2 and C5. Besides, the rank constraint in C6 is combinatorial.We first handle constraints C2 and C5 by transforming them  into linear matrix inequalities (LMIs) using the following lemma:

\begin{Lem}[S-Procedure \cite{book:convex}] \label{lem:S_Procedure} Let a function $f_m(\mathbf{x}),m\in\{1,2\},\mathbf{x}\in \mathbb{C}^{N\times 1},$ be defined as
\begin{eqnarray}\label{eqn:S-procedure}
f_m(\mathbf{x})=\mathbf{x}^H\mathbf{A}_m\mathbf{x}+2 \mathrm{Re} \{\mathbf{b}_m^H\mathbf{x}\}+c_m,
\end{eqnarray}
where $\mathbf{A}_m\in\mathbb{H}^N$, $\mathbf{b}_m\in\mathbb{C}^{N\times 1}$, and $c_m\in\mathbb{R}$. Then, the implication $f_1(\mathbf{x})\le 0\Rightarrow f_2(\mathbf{x})\le 0$  holds if and only if there exists a $\delta\ge 0$ such that
\begin{eqnarray}\delta
\begin{bmatrix}
       \mathbf{A}_1 & \mathbf{b}_1          \\
       \mathbf{b}_1^H & c_1           \\
           \end{bmatrix} -\begin{bmatrix}
       \mathbf{A}_2 & \mathbf{b}_2          \\
       \mathbf{b}_2^H & c_2           \\
           \end{bmatrix}          \succeq \zero,
\end{eqnarray}
provided that there exists a point $\mathbf{\hat{x}}$ such that $f_m(\mathbf{\hat{x}})<0$.
\end{Lem}

 Applying Lemma 1,  the original constraint C2 holds if and only if there exists a $\delta_k\ge 0$,  such that the following  LMI constraint holds:
\begin{eqnarray}\label{eqn:LMI_C2}
\mbox{C2: }\mathbf{S}_{\mathrm{C}_{2_k}}\Big(\mathbf{W}_k,\delta_k\Big)&=&
          \begin{bmatrix}
       \delta_k\mathbf{I}_{N_\mathrm{T}}  & \hspace*{-1mm}\zero        \\
       \zero     & \hspace*{-1mm}-\delta_k\rho^2_k -\sigma_\mathrm{s}^2     \\
           \end{bmatrix}\\
           & +&\mathbf{U}_{\mathbf{\hat h}_k}^H\Big(\frac{\mathbf{W}_k}{\Gamma_{\mathrm{req}_k}}-\sum_{i\neq k}\mathbf{W}_i\Big)\mathbf{U}_{\mathbf{\hat h}_k} \succeq \mathbf{0},\notag
\end{eqnarray}
where $\mathbf{U}_{\mathbf{\hat h}_k}=\Big[\mathbf{I}_{N_{\mathrm{T}}}\quad\mathbf{\hat h}_k\Big]$. Similarly,  constraint C5 can be equivalently written as
\begin{eqnarray}\label{eqn:LMI_C5}
&&\hspace*{-16mm}\mbox{C5: }\mathbf{S}_{\mathrm{C}_{5_j}}\Big(\mathbf{W}_k,\mathbf{W}_{\mathrm{E}}, \bm{\nu},\bm{\beta}\Big)\\
&\hspace*{-16mm}=&\notag
           \begin{bmatrix}
       \nu_j\mathbf{I}_{N_{\mathrm{T}}N_{\mathrm{R}}}& \zero          \\
        \zero
        & -\beta_j-\nu_j\upsilon_j^2       \\
           \end{bmatrix}\notag\\
           &\hspace*{-16mm}+& \mathbf{U}_{\widetilde{\mathbf{g}}_j}^H\Bigg\{\sum_{k=1}^K(\mathbfcal{ W}_k+\mathbfcal{W}_{\mathrm{E}})\Bigg\}\mathbf{U}_{\widetilde{\mathbf{g}}_j}\succeq \mathbf{0}, \forall j,\notag
\end{eqnarray}
where $\bm\nu=\{ \nu_1,\ldots,\nu_j,\ldots,\nu_J\}$,  $\nu_j\ge 0$, $\mathbfcal{ W}_k=\mathbf{I}_{N_\mathrm{R}} \otimes \mathbf{W}_k $, $\mathbfcal{W}_{\mathrm{E}}=\mathbf{I}_{N_\mathrm{R}} \otimes \mathbf{W}_{\mathrm{E}} $, $\mathbf{U}_{\widetilde{\mathbf{g}}_j}=[\mathbf{I}_{N_{\mathrm{T}}N_{\mathrm{R}}}\quad \widetilde{\mathbf{g}}_j]$, and $\widetilde{\mathbf{g}}_j=\vect({\mathbf{\hat G}}_j)$. Then, the considered optimization problem can be rewritten as
\begin{eqnarray}\label{eqn:rank-constrained-SDP-2}
\underset{\underset{\bm{\beta},\tau}{     \mathbf{W}_\mathrm{E},\mathbf{W}_k\in \mathbb{H}^{N_{\mathrm{T}}},   }}{\maxo}\,\, \hspace*{-14mm}&& \hspace*{14mm}  \tau\\
\hspace*{-14mm} \mathrm{s.t.}\,\, &&\hspace*{-6mm}\mathrm{C1, C3, C4, C7,}\notag\\
\hspace*{-14mm} &&\hspace*{-6mm}\mathrm{C2}:\,\, \mathbf{S}_{\mathrm{C}_{2_k}}\Big(\mathbf{W}_k,\delta_k\Big) \succeq \zero, \forall k,\notag\\
\hspace*{-14mm} &&\hspace*{-6mm}\mathrm{C5}:\,\, \mathbf{S}_{\mathrm{C}_{5_j}}\Big(\mathbf{W}_k,\mathbf{W}_{\mathrm{E}}, \bm{\nu},\bm{\beta}\Big)\succeq \zero,\forall j,\notag\\
\hspace*{-14mm} &&\hspace*{-6mm}\mathrm{C6}:\,\, \Rank(\mathbf{W}_k)\le 1,\notag
\end{eqnarray}
\noindent
where  $\delta_k\ge 0$ and $\bm\nu\ge0$  are the auxiliary optimization variables introduced in Lemma 1 for handling constraints C2 and C5, respectively.  We note that C2 and C5 are now LMIs with finite numbers of constraints which are relatively easier to handle compared to the infinite numbers of constraints in the original problem formulation. However, the rank constraint in C6 is still an obstacle to solving the considered optimization problem due to its combinatorial nature. As a result, we adopt SDP relaxation by removing constraint C6 from the problem formulation which yields:
\begin{eqnarray}\label{eqn:rank-constrained-SDP-2}
\underset{\underset{\bm{\beta},\tau}{     \mathbf{W}_\mathrm{E},\mathbf{W}_k\in \mathbb{H}^{N_{\mathrm{T}}},   }}{\maxo}\,\, \hspace*{-14mm}&& \hspace*{14mm}  \tau\\
\hspace*{-14mm} \mathrm{s.t.}\,\, &&\hspace*{-6mm}\mathrm{C1, C3, C4, C7,}\notag\\
\hspace*{-14mm} &&\hspace*{-6mm}\mathrm{C2}:\,\, \mathbf{S}_{\mathrm{C}_{2_k}}\Big(\mathbf{W}_k,\delta_k\Big) \succeq \zero, \forall k,\notag\\
\hspace*{-14mm} &&\hspace*{-6mm}\mathrm{C5}:\,\, \mathbf{S}_{\mathrm{C}_{5_j}}\Big(\mathbf{W}_k,\mathbf{W}_{\mathrm{E}}, \bm{\nu},\bm{\beta}\Big)\succeq \zero,\forall j,\notag\\
\hspace*{-14mm} &&\hspace*{-6mm}\cancel{\mathrm{C6}:\,\, \Rank(\mathbf{W}_k)\le 1}.\notag
\end{eqnarray}
\noindent

The problem in \eqref{eqn:rank-constrained-SDP-2} is a standard convex optimization problem and can be solved numerically with computationally efficient off-the-shelf  convex programs solvers such as CVX \cite{website:CVX}. However, it is unclear if the obtained solution satisfies $\Rank(\mathbf{W}_k)\leq 1,\forall k$. Therefore, we introduce the following theorem to reveal the structure of the solution of \eqref{eqn:rank-constrained-SDP-2}.

\begin{Thm}\label{thm:rankone}
Let the optimal beamforming matrix and energy covariance matrix  of \eqref{eqn:rank-constrained-SDP-2} be $\mathbf{W}_k^*$ and $\mathbf{W}_{\mathrm{E}}^*$, respectively. Assuming the considered problem is feasible for $P_{\max}>0$ and $\Gamma_{\mathrm{req}_k}>0,\forall k$, then $\Rank(\mathbf{W}_k^*)=1,\forall k$, and $\Rank(\mathbf{W}_{\mathrm{E}}^*)\leq 1$.
\end{Thm}

\,\,\emph{Proof:} Please refer to the Appendix.

Thus, the globally optimal solution of \eqref{eqn:rank-constrained-SDP-2} can be obtained. In particular, employing information beamforming for each IR and at most one energy beam is optimal for the considered problem, despite the imperfection of the CSI and the non-linearity of the RF EH circuits.

\section{Simulation}\label{sect:simulation}
In this section, the performance of the proposed optimal beamforming design is evaluated via simulations. The important simulation parameters are summarized in Table \ref{table:parameters}.
Unless specified otherwise, we assume that there are $K=2$ IRs and $J=4$ ERs which are located $100$ meters and $5$ meters from the transmitter, respectively.  Furthermore,  we choose the normalized maximum  channel estimation errors of ER $j$  and IR $k$ as  $\sigma_{\mathrm{est}_{G_j}}^2=1\%\ge\frac{\upsilon^2_j}{\norm{\mathbf{G}_j}^2_F},\forall j,$  and $\sigma_{\mathrm{est}_{h_k}}^2=1\%\ge\frac{\rho^2}{\norm{\mathbf{h}_k}^2_2}$, respectively. Besides, we assume that all IRs require the same minimum data rate, i.e., $\Gamma_{\mathrm{req}_k}=\Gamma_{\mathrm{req}}$. For the non-linear EH circuits, we set the maximum harvested power per wireless powered device to $M_j=24$ mW. Besides, we adopt $a_j=150$ and $b_j=0.014$.  We solve the
optimization problem in \eqref{eqn:TP_maximization} and obtain the average system
performance by averaging over different channel realizations.

\begin{table}[t]
\caption{Simulation Parameters.} \label{table:parameters}

\begin{tabular}{ | l | l | }\hline
      Carrier center frequency                           & $
      915$ MHz\\ \hline
      Bandwidth                                          & $200$ kHz \\ \hline 
      Transceiver  antenna gain                                     & $10$ dBi \\ \hline
      Number of receive antennas  $N_\mathrm{R}$                                  & $2$ \\ \hline
      Noise power              $\sigma^2$                          & $ -95$ dBm \\ \hline
          Maximum transmit power    $P_{\max}$                                  & $36$ dBm \\ \hline
    Transmitter-to-ER fading distribution                                   & Rician with Rician factor $3$ dB \\
        \hline
         Transmitter-to-IR fading distribution     \quad                            & Rayleigh \\
\hline
\end{tabular}
\end{table}

\begin{figure}[t]
        \centering
       \includegraphics[width=3.5 in]{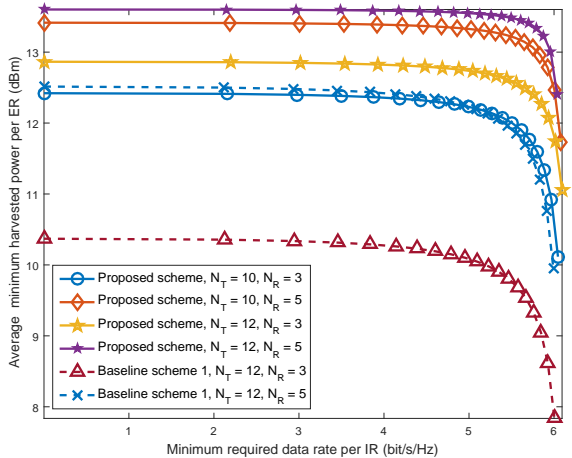}
        \caption{ Average minimum harvested power  per ER (dBm) versus the minimum required data rate per IR (bit/s/Hz) for different numbers of antennas.}\label{fig:TP_energy_trade_off}
        \end{figure}

\begin{figure}[t]        \centering
        \includegraphics[width=3.5 in]{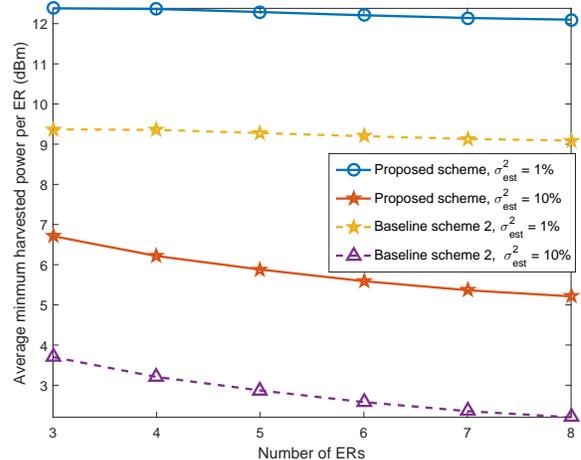}
        \caption{Average minimum harvested power per ER (dBm) versus the number of ERs for $N_{\mathrm{T}}=10$ and $N_{\mathrm{R}}=3$.}
        \label{fig:TP_ERs}
\end{figure}

In Figure \ref{fig:TP_energy_trade_off}, we show the average minimum harvested power per ER in the considered SWIPT systems for the optimal robust beamforming design. As can be observed, there is a non-trivial trade-off between  the minimum harvested power per ER and the minimum required data rate per IR. In particular,  the achievable data rate per user and the minimum harvested power per ER cannot be maximized simultaneously. Besides, for the optimal resource allocation, the trade-off region of the minimum achievable rate and the harvested energy is enlarged significantly  for larger $N_{\mathrm{T}}$ and $N_{\mathrm{R}}$. This is due to the fact that the extra degrees of freedom offered by multiple transmit antennas can be used to focus both the information and energy beams which improves the beamforming  efficiency.  On the other hand, increasing the number of receive antennas $N_{\mathrm{R}}$ at the ER can significantly improve the minimum harvested energy per ER. In fact, the additional receive antennas act as additional energy collectors  which enable a more efficient energy transfer. Furthermore, we verified by simulation that $\Rank(\mathbf{W}_k)=1$ can be obtained for all considered channel realizations which confirms the correctness of Theorem \ref{thm:rankone}. On the other hand, we also show  the performance of a baseline scheme for comparison in Figure \ref{fig:TP_energy_trade_off}. For baseline scheme 1,  an existing linear EH model with $\eta_j=1$, cf. \eqref{eqn:linear_EH_model}, is adopted for resource allocation algorithm design. Specifically, we optimize $\mathbf{w}_k$ and $\mathbf{W}_{\mathrm{E}}$ to maximize the minimum harvested power  per ER  subject to the constraints in \eqref{eqn:TP_maximization}. Then, the resource allocation designed by  baseline scheme 1 is applied in the considered system with non-linear ERs.  We observe from Figure \ref{fig:TP_ERs} that
 a substantial gain in harvested power is achieved by the proposed optimal resource allocation algorithm compared to baseline scheme 1. This is because baseline scheme 1 does not take into account the non-linearity of practical EH circuits leading to mismatches in resource allocation.

In Figure \ref{fig:TP_ERs}, we study the average minimum harvested power per ER versus the number of ERs  for different maximum normalized channel estimation error variances and beamforming schemes. The minimum required SINR is set to $\Gamma_{\mathrm{req}_k}=10$ dB and there are $N_\mathrm{R}=3$ and $N_\mathrm{T}=10$ antennas equipped at each ER and the transmitter, respectively. Besides, the maximum normalized channel estimation error variance of the transmitter-to-IR links and  the transmitter-to-ER links  are set to be identical, i.e.,  $\sigma_{\mathrm{est}_{G_j}}^2=\sigma_{\mathrm{est}_{h_k}}^2=\sigma_{\mathrm{est}}^2$. As can be observed, the average minimum harvested power per ER in the system decreases with an increasing number of ERs. In fact, the more ERs are in the system, the more difficult it is for the transmitter to provide fair resource allocation for all ERs.     In particular, for a large number of ERs in the system, it is more likely that there are some ERs with poor channel qualities. Thus, the transmitter is forced to steer the information and energy signals toward ERs with weak channel conditions which reduces the minimum harvested power per ER.  On the other hand, the average minimum harvested power per ER decreases  with increasing $\sigma_{\mathrm{est}}^2$, since the CSI quality degrades with increasing $\sigma_{\mathrm{est}}^2$. In particular, for a larger value of $\sigma_{\mathrm{est}}^2$,  it becomes more difficult for the transmitter  to accurately focus the transmitter energy as would be necessary for achieving a high efficiency in SWIPT. For comparison, we also show the performance of baseline scheme 2 which adopts an isotropic radiation pattern for $\mathbf{W}_{\mathrm{E}}$. Then, we maximize the minimum harvested power per ER by optimizing $\mathbf{W}_k$ and the power of $\mathbf{W}_{\mathrm{E}}$ subject  to  the
same  constraints as  in \eqref{eqn:Rank-constrained} via SDP relaxation. It can be seen that the performance of the baseline scheme is unsatisfactory compared to the proposed scheme. In fact, baseline scheme 2  cannot fully exploit the available degrees of freedom for efficient WPT as the beamforming direction of $\mathbf{W}_{\mathrm{E}}$ is fixed.

\section{Conclusions}\label{sect:conclusion}
We studied the beamforming design for multiuser SWIPT systems with the objective of ensuring max-min fairness in WPT.
The  design was formulated as a non-convex optimization problem which took into account the  non-linearity of practical EH circuits  and the imperfection of the CSI. The optimization problem was solved by applying SDP relaxation.  Simulation results demonstrated that the proposed optimal beamforming design offers significant performance gains compared to two baseline schemes.

\section*{Appendix-Proof of Theorem \ref{thm:rankone}}\label{app:rankone}
We follow a similar approach as in \cite{JR:Elena_TCOM} to prove Theorem 1. We note that the relaxed problem in \eqref{eqn:rank-constrained-SDP-2} is jointly convex with respect to the optimization variables. Besides, it can be verified that the problem satisfies the Slater's constraint qualification and thus has a zero duality gap. Therefore, to reveal the structure of $\mathbf{W}_k$ and $\mathbf{W}_{\mathrm{E}}$, we consider the Lagrangian of problem \eqref{eqn:rank-constrained-SDP-2} which is given by:
\begin{eqnarray}
L\hspace*{-3mm}&=&\hspace*{-3mm}\tau-\lambda\Big(\sum_{k=1}^K\Tr(\mathbf{W}_k)+\Tr(\mathbf{W}_{\mathrm{E}})- P_{\max}\Big) \\
\hspace*{-3mm}&+&\hspace*{-3mm}\sum_{k=1}^K\Tr(\mathbf{S}_{\mathrm{C}_{2_k}}\Big(\mathbf{W}_k,\delta_k\Big)\mathbf{D}_{\mathrm{C}_{2_k}})+ \sum_{k=1}^K\Tr(\mathbf{W}_k\mathbf{Y}_k)\notag\\
\hspace*{-3mm}&+&\hspace*{-3mm}\sum_{j=1}^J\Tr(\mathbf{S}_{\mathrm{C}_{5_j}}\Big(\mathbf{W}_k,\mathbf{W}_{\mathrm{E}}, \bm{\nu},\bm{\beta}\Big)\mathbf{D}_{\mathrm{C}_{5_j}})+ \Tr(\mathbf{W}_{\mathrm{E}}\mathbf{Z})+{\bm\Delta},\notag
\end{eqnarray}
where $\lambda\geq 0$, $\mathbf{D}_{\mathrm{C}_{2_k}}\succeq\zero$,  $\mathbf{Z}\succeq\zero$, $\mathbf{D}_{\mathrm{C}_{5_j}}\succeq\zero,\forall j\in\{1,\ldots,J\}$, and $\mathbf{Y}_k\succeq\zero$  are the dual variables for constraints C1, C2, C3, C5, and C7,  respectively.  Besides, we denote $\bm\Delta$ as a collection of variables and constants that are not relevant to the proof. For notational convenience, we denote the optimal
primal and dual variables of the SDP relaxed version of \eqref{eqn:rank-constrained-SDP-2}  by the corresponding
variables with an asterisk superscript in the following. Now, we focus on those Karush-Kuhn-Tucker (KKT) conditions which are needed for the proof.

\begin{subequations}
\begin{eqnarray}
&&\mathbf{Y}^*_k,\mathbf{Z}^*,\mathbf{D}_{\mathrm{C}_{3_k}}^*,\mathbf{D}_{\mathrm{C}_{5_j}}^*\succeq \zero,\quad\lambda^*\ge0,\\
&& \mathbf{Y}_k^*\mathbf{W}_k^*=\zero,\quad \label{eqn:KKT-complementarity}   \mathbf{Z}^*\mathbf{W}_{\mathrm{E}}^*=\zero, \label{eqn:KKT-complementarity}\\
\label{eqn:KKT_Y1}
&&\mathbf{Y}^*_k=\lambda^*\mathbf{I}_{N_\mathrm{T}}-{\bm\Xi}_k, \\
&& {\bm\Xi}_k=\mathbf{U}_{\mathbf{\hat h}_k}\frac{\mathbf{D}^*_{\mathrm{C}_{2_k}}}{\Gamma_{\mathrm{req}_k}}\mathbf{U}_{\mathbf{\hat h}_k}^H- \sum_{i\ne k}\mathbf{U}_{\mathbf{\hat h}_i}{\mathbf{D}^*_{\mathrm{C}_{2_i}}}\mathbf{U}_{\mathbf{\hat h}_i}^H\notag\\
&&+\sum_{j=1}^J\sum_{l=1}^{N_{\mathrm{R}}} \Big[
\mathbf{U}_{\widetilde{\mathbf{g}}_j}\mathbf{D}^*_{\mathrm{C}_{5_j}}\mathbf{U}_{\widetilde{\mathbf{g}}_j}^H \Big]_{a:b,c:d}, \\
\label{eqn:KKT_Y2}&&\mathbf{Z^*}= \lambda^*\mathbf{I}_{N_\mathrm{T}}-\sum_{j=1}^J\sum_{l=1}^{N_{\mathrm{R}}} \Big[
\mathbf{U}_{\widetilde{\mathbf{g}}_j}\mathbf{D}^*_{\mathrm{C}_{5_j}}\mathbf{U}_{\widetilde{\mathbf{g}}_j}^H \Big]_{a:b,c:d}, \end{eqnarray}
\end{subequations}\noindent
where $ a=(l-1)N_{\mathrm{T}}+1,b=l N_{\mathrm{T}},c=(l-1)N_{\mathrm{T}}+1,$ and $d=l N_{\mathrm{T}}$.

From \eqref{eqn:KKT-complementarity}, we know that the columns of  $\mathbf{W}_k^*$ lie in the null space of $\mathbf{Y}^*_k$. In order to reveal the rank of $\mathbf{W}_k^*$, we investigate the structure of $\mathbf{Y}^*_k$. First, it can be shown that $\lambda^*>0$ since constraint C1 is active at the optimal solution. Then, we show that  $\bm{\Xi}_k$  in \eqref{eqn:KKT_Y1} is a positive semidefinite matrix by contradiction. Suppose  $\bm{\Xi}_k$ is a negative definite matrix, then from  \eqref{eqn:KKT_Y1},  $\mathbf{Y}_{k}^{*}$ becomes a full-rank and positive definite matrix. By \eqref{eqn:KKT-complementarity}, $\mathbf{W}_k^*$ is forced to be the zero matrix which is not an optimal solution for $P_{\max}>0$ and $\Gamma_{\mathrm{req}_k}>0$.  Thus, in the following, we focus on the case  ${\bm \Xi}_k\succeq \zero$. Since matrix $\mathbf{Y}^*_k=\lambda^*\mathbf{I}_{N_\mathrm{T}}-{\bm\Xi}_k$ is positive semidefinite, the following inequality holds: \begin{eqnarray}
\lambda^* \geq \lambda_{\mathbf{\Xi}_k}^{\max} &\ge& 0,
\end{eqnarray}
where $\lambda_{\mathbf{\Xi}_k}^{\max}$ is the maximum eigenvalue of matrix $\bm{\Xi}_k$. From \eqref{eqn:KKT_Y1},  if $\lambda^*> \lambda_{\mathbf{\Xi}_k}^{\max} $, matrix $\mathbf{Y}_{k}^{*}$ will become a positive definite matrix with full rank. However, this will again yield the solution $\mathbf{W}_{k}^{*} = \zero$ which is not optimal since $\Gamma_{\mathrm{req}}>0$. Thus, at  the optimal solution, $\lambda^* = \lambda_{\mathbf{\Xi}_k}^{\max}$ must holds. Besides, in order to have a bounded optimal dual solution, it follows that the null space of $\mathbf{Y}_{k}^{*}$ is spanned by vector $\mathbf{u}_{\mathbf{\Xi}_k,\max}\in\mathbb{C}^{N_\mathrm{T}\times1}$, which is the unit-norm eigenvector of $\mathbf{\Xi}_k$ associated with eigenvalue $\lambda_{\mathbf{\Xi}_k}^{\max}$. As a result, the optimal beamforming matrix $\mathbf{W}_k^*$ has to be a rank-one matrix and is given by
\begin{equation}
\mathbf{W}_k^* = \psi \mathbf{u}_{\mathbf{\Xi}_k,\max} \mathbf{u}_{\mathbf{\Xi}_k,\max}^H.
\end{equation}
where $\psi$ is a parameter such that the power consumption satisfies constraint C2.

On the other hand, for revealing the structure of $\mathbf{Z}^*$, we focus on \eqref{eqn:KKT_Y2}. Define an auxiliary variable matrix $\mathbf{B}=\sum_{j=1}^J\sum_{l=1}^{N_{\mathrm{R}}} \Big[
\mathbf{U}_{\widetilde{\mathbf{g}}_j}\mathbf{D}^*_{\mathrm{C}_{5_j}}\mathbf{U}_{\widetilde{\mathbf{g}}_j}^H \Big]_{a:b,c:d}\succeq \zero$ and the corresponding  maximum eigenvalue $\lambda_{\mathbf{B}}^{\max}$. Since $\mathbf{Z}^*\succeq \zero$, we have $\lambda^*\ge \lambda_{\mathbf{B}}^{\max}\geq 0$. If $\lambda^*= \lambda_{\mathbf{B}}^{\max}$, then  $\Rank(\mathbf{Z}^*)=N_{\mathrm{T}}-1 $ and  $\Rank(\mathbf{W}_{\mathrm{E}}^*)=1$. If $\lambda^*> \lambda_{\mathbf{B}}^{\max}$,
then  $\Rank(\mathbf{Z}^*)=N_{\mathrm{T}} $ and  $\Rank(\mathbf{W}_{\mathrm{E}}^*)=0$. Therefore,   $\Rank(\mathbf{W}_{\mathrm{E}}^*)\leq 1$ and at most one  energy beam is required to achieve optimality.
\hfill\qed


\end{document}